\documentclass[
reprint,
nofootinbib,
amsmath,amssymb,amsbsy,
prd,
twocolumn,
superscriptaddress,
longbibliography,
preprintnumbers]{revtex4-1}
\usepackage[dvipsnames,usenames]{xcolor}
\usepackage{graphicx}
\usepackage{subfig}
\usepackage{dcolumn}
\usepackage{bm}
\usepackage{mathrsfs,natbib}
\usepackage[title]{appendix}
\usepackage{graphicx,epsf,amssymb,amsbsy,amsfonts,amssymb,amsmath}
\usepackage[colorlinks=true]{hyperref}
\hypersetup{
    unicode=false,          
    pdftoolbar=true,        
    pdfmenubar=true,        
    pdffitwindow=false,     
    pdfstartview={FitH},    
    pdftitle={},    
    pdfauthor={},     
    pdfsubject={},   
    pdfcreator={},   
    pdfproducer={}, 
    pdfkeywords={} {} {}, 
    pdfnewwindow=true,      
    colorlinks=true,       
    linkcolor=magenta, 
    citecolor=blue,        
    filecolor=magenta,      
    urlcolor=blue           
}
\usepackage{slashed}
\usepackage{comment}
\usepackage[dvipsnames]{xcolor}
\usepackage{color,soul}
\usepackage{simplewick}

\newcommand\R{\mathcal{R}}

\newcommand{\be}{\begin{equation}}
\newcommand{\ee}{\end{equation}}
\newcommand\beq{\begin{eqnarray}}
\newcommand\eeq{\end{eqnarray}}

\newcommand{\mybar}[1]

\newlength{\backup}

\begin{document}

\title{Gauge field flow for chiral gauge theories on a disk boundary}
\author{Jinlong Dang}%
\email{dangjl@stu.pku.edu.cn}
\affiliation{School of Physics, Peking University, Beijing 100871, China}
\author{Rohith Karur}%
\email{r\_karur137@berkeley.edu}
\affiliation{Department of Physics, University of California, Berkeley, CA 94720, USA}
\affiliation{Nuclear Science Division, 
Lawrence Berkeley National Laboratory, Berkeley, 
CA 94720, USA}
\author{Srimoyee Sen}%
\email{srimoyee08@gmail.com}
\affiliation{Department of Physics and Astronomy, Iowa State University, Ames, Iowa 50011, USA}%

\date{\today}

\begin{abstract}
   
A recent non-perturbative formulation of $2n$ dimensional  chiral gauge theories relies on realizing chiral fermions on the $2n$ dimensional boundary of a $2n+1$ dimensional disk manifold \cite{Kaplan:2023pxd, Kaplan:2023pvd}. It also requires extending boundary gauge configurations into the interior of the disk using some flow prescription that preserves 2n dimensional gauge invariance. In this paper we propose a concrete realization of the equation of motion flow with the disk embedded on a square lattice. In addition, we couple the flow gauge field to fermions and demonstrate the mechanism of anomaly inflow and anomaly cancellation at work on the lattice.
\end{abstract}
\maketitle
\section{Introduction}
Non-perturbative formulations of quantum field theories are essential not only for the extraction of physical observables, but also for uncovering the underlying structure of the theories themselves. Lattice regularization provides the most systematic framework for such a definition. While it has been remarkably successful in addressing the non-perturbative dynamics of quantum chromodynamics, extending this success to the full Standard Model has remained an outstanding challenge. The difficulty lies in constructing a lattice regulator that reproduces the correct continuum limit in the presence of chiral gauge interactions. In the Standard Model, the weak interactions couple asymmetrically to left- and right-handed fermions, placing it within the broader class of chiral gauge theories. Despite decades of effort, a general and fully satisfactory non-perturbative formulation of chiral gauge theories has remained elusive.  

The obstruction can be traced back to the Nielsen-Ninomiya theorem \cite{Nielsen:1980rz, Nielsen:1981xu, Nielsen:1981hk, Karsten:1981gd, Friedan:1982nk} or the fermion doubling problem. One of the early attempts at overcoming this problem was based on the idea of domain wall fermions \cite{Kaplan:1992bt} where one aims to utilize a microscopic theory defined in $2n+1$ dimensions to recover $2n$ dimensional chiral gauge theories as a low energy effective field theory. The idea however proved to be useful for vector gauge theories like quantum chromodynamics(QCD) as opposed to chiral gauge theories. Another attempt to getting around the Nielsen Ninomiya theorem which goes by the name of overlap fermions \cite{Neuberger:1997fp, Narayanan:1993sk}, too, was found to be suitable for vector gauge theories like QCD. In fact, domain wall fermions and overlap fermions were found to have deep conceptual ties \cite{Neuberger:1997bg}. Eventually, in \cite{Luscher:1998du} the overlap operator was used to provide a non-perturbative formulation for Abelian chiral gauge theories (CGT). However, its non-Abelian extension has remained out of reach since. We refer to these approaches as extra dimension based approaches. Several other directions have been pursued over the years to findd a non-perturbative formulation of CGTs. Among these, one of the most promising is symmetric mass generation (SMG). Some of the earliest work in this direction are \cite{Eichten:1985ft, Golterman:1992yha}. SMG based proposals have seen a recent revival \cite{Wen:2013ppa, You:2014sqa, You:2014vea, Wang:2018ugf, Catterall:2020fep, Wang:2022ucy, Razamat:2020kyf} owing to \cite{PhysRevB.83.075103}. In addition, there have been bosonization based approaches \cite{Berkowitz:2023pnz}, symmetry entangler based approach \cite{Thorngren:2026ydw}, anomaly mediated super-symmetry breaking approaches \cite{Leedom:2025mcg, Csaki:2021xhi, Csaki:2021aqv}, functional renormalization group approach \cite{Li:2025tvu, Li:2026ayh} and gauge fixing approach \cite{Kadoh:2007wz, Kikukawa:2017ngf, Bock:1997ks}.  

This paper concerns itself with the extra dimension or domain wall fermion (DWF) based approach, where recent breakthroughs are indicating a new path forward \cite{Kaplan:2023pxd, Kaplan:2023pvd, Kaplan:2024ezz}. The difficulty of using the original DWF based approach can be summarized as follows. The DWF was originally formulated on slab geometry where a domain wall is always accompanied with an anti-wall, which 
necessarily gives rise to Weyl fermions in pairs of opposite chirality with wall and the anti-wall hosting opposite chiralities. The anti-wall chiral mode is usually called the mirror fermion.
The gauge fields are taken to be extra dimension independent which couples identically to the two chiralities 
leading to a vector-like theory. In 2016, \cite{Grabowska:2015qpk} attempted to remedy this while retaining the slab geometry, but forcing the gauge field coupling to fermions to decay as a function of the extra dimension. The hope was to decouple the mirror (anti-wall) fermions from the gauge fields, while retaining the gauge coupling of the fermions on the wall. A recent breakthrough went even further \cite{Kaplan:2023pxd} and modified the geometry of the construction in such a way so as to engineer a single wall without an anti-wall, which in turn helped realize for the first time an unpaired Weyl fermions on a finite lattice \cite{Kaplan:2023pvd}. In other words, the construction of \cite{Kaplan:2023pvd} eliminated the mirrors fermions modes. The main idea here is to begin with a $2n+1$ dimensional ($\mathbb{R}_2\times \mathbb{T}_{2n-1}$) Dirac fermion with a mass defect with the mass changing sign as a function of the radial coordinate in $\mathbb{R}_2$. This gives rise to a single unpaired Weyl fermion localized on the defect where the mass crosses zero. To complete the formulation, one needs to couple the fermion to $2n$-dimensional gauge fields and then extend this gauge field into the extra dimension in the interior of the disk in a way that preserves $2n$ dimensional gauge invariance. \cite{Kaplan:2024ezz} proposed using $2n+1$-dimensional equation of motion to prescribe  the gauge fields in the interior of the disk. This process is reminiscent of lattice gradient flow and \cite{Kaplan:2024ezz} referred to it as EOM flow. The goal of the present work is to formulate this EOM flow on a Euclidean square lattice and to investigate its properties in a controlled setting. In particular, we demonstrate the implementation of the flow for a specific class of $2n$-dimensional gauge field configurations and use it to illustrate the mechanisms of anomaly inflow and anomaly cancellation in the resulting theory. For concreteness, we focus on the case of $n=1$ with Abelian gauge fields, where the essential features can be made explicit. The generalization to higher dimensions and non-Abelian gauge theories is expected to follow straightforwardly. Note that \cite{PaperA} formulated the lattice gradient flow proposed in \cite{Grabowska:2015qpk}. 

The main contributions of this paper are three fold:
\begin{itemize}
\item Noticing that equation of motion by itself does not lead to a unique flow prescription. An additional condition is required for this purpose. In this paper, we set this additional condition by demanding that the gauge field configuration be a stable local minimum of the higher dimensional gauge field action whose extremization leads to the equation of motion. 
\item  Introducing an additional imaginary time direction and using gradient flow in this additional time to arrive at the stable local minimum of the higher dimensional gauge field action. 

\item In addition, our work provides a pathway to constructing a radial flow, on a square lattice. In the process of doing so, we address subtleties that arise from the fact that a radial flow is most naturally realized in a polar coordinate system which however is not ideal for a square lattice. 
\end{itemize}

The organization of the paper is as follows. We begin with an overview of the physics in continuous space-time in sec. \ref{sec:ov}. In sec. \ref{sec:latimpl} we formulate the construction of the flow on a square lattice. Subsequently, in section \ref{sec:ferm} we obtain the flow gauge field for a specific boundary gauge field configuration and couple it to fermions, compute current densities and demonstrate lattice anomaly inflow and anomaly cancellation at work.

\section{Overview}
\label{sec:ov}
In this section we present the main ideas which are best explained in the continuum. In the following sections we will translate them to a square lattice.
The ingredients we need to construct $2n$ dimensional chiral gauge theories are (i) $2n$ dimensional Weyl fermions of appropriate chirality and (ii) gauge fields that interact with these Weyl fermions in the requisite representations. The chiral gauge theories of practical interest pertain to $n=2$. However, for the purpose of illustration we will work in $n=1$. The prototypical model of chiral gauge theories in $n=1$ or $1+1$ dimensions, is the $3-4-5-0$ model where one encounters two Weyl fermions of positive or right chirality and two Weyl fermions of negative or left chirality interacting with a $U(1)$ gauge field: the charge of the two negative or left chirality fermions being $+3, +4$ and the two right or positive chirality fermions being $5, 0$. One could of course reverse the chirality of each fermion and the resulting theory would still be a chiral gauge theory.  

Engineering this model using the extra dimension framework proposed in \cite{Kaplan:2023pvd} can be broken down into two steps. The first is to construct the Weyl fermion contents of the target theory, the second is to construct the gauge field sector. The construction of both left and right chirality Weyl fermions using the extra dimension approach was outlined in \cite{Kaplan:2023pxd, Kaplan:2023pvd} and we follow the same procedure here. Note that, we will not implement Monte-Carlo sampling of gauge configurations for calculating observables. Instead the focus of this work is: (i) formulate the gauge field flow in the extra dimension given any boundary gauge field configuration (ii) for a specific boundary gauge field, demonstrate the anomaly inflow mechanism for a specific chirality of Weyl fermion with charge $1$ which is then followed by a demonstration of anomaly cancellation on the lattice for the $3-4-5-0$ model. 

Following \cite{Kaplan:2023pxd, Kaplan:2023pvd}, to engineer a left chirality fermion with the extra dimension setup, one begins with a $2+1$ dimensional manifold $\mathbb{R}_2\times \mathbb{T}_1$ and introduces $2+1$ dimensional Dirac fermion with a position dependent mass where using polar coordinates in $\mathbb{R}_2$: $r, \phi$ 
\beq
m(r)=\begin{cases}
m_0 \,\,\,\,\text{for}\,\, r<R\\
-m_0\,\,\,\,\text{for}\,\, r>R
\end{cases}
\label{mr}
\eeq
with $m_0>0$, can in fact the cutoff scale. We will denote the coordinate corresponding to $\mathbb{T}_1$ as $t$. Also, in this 3d Euclidean space, we can choose the following representation of gamma matrices:
\beq
    \gamma_t &=& \sigma_3 = \begin{pmatrix} 1 & 0 \\ 0 & -1 \end{pmatrix}, \quad
    \gamma_s = \sigma_1 = \begin{pmatrix} 0 & 1 \\ 1 & 0 \end{pmatrix},\nonumber\\
    \gamma_x &=& \sigma_2 = \begin{pmatrix} 0 & -i \\ i & 0 \end{pmatrix}
    \label{gm}
\eeq
where $r=\sqrt{s^2+x^2}$ and $\phi=\tan^{-1}(x/s)$. We also define
\beq
    \gamma_r &=& \gamma_s \cos\phi + \gamma_x \sin\phi = 
    \begin{pmatrix}
        0 & e^{-i\phi}\\
        e^{i\phi} & 0 
    \end{pmatrix}\nonumber\\ 
    \gamma_\phi &=& -\gamma_s \sin\phi + \gamma_x\cos\phi = 
    \begin{pmatrix}
        0 & -i e^{-i\phi} \\
        ie^{i\phi} & 0 
    \end{pmatrix}
    \label{gm2}
\eeq
following the convention of ref.\cite{Kaplan:2023pxd}. 
The long wavelength excitations of this system correspond to chiral fermion modes located at $r=R$ \cite{Kaplan:2023pxd}. Note that, $\gamma_r$ becomes the chirality matrix for the lower dimensional fermions localized on $r=R$.  
For the specific choice of gamma matrices and the mass defect of Eq. \ref{gm}, \ref{gm2}, the modes at $r=R$ are left handed or of negative chirality, i.e. are eigenstates of the operator $\gamma_r$ with eigenvalues of $-1$. To engineer a right chirality fermion, one simply reverses the sign of the mass defect, i.e. taking the sign of the mass to be negative for $r<R$ and positive for $r>R$. 

We are to now introduce gauge field in this construction. The target theory of interest involves $2$ dimensional dynamical $U(1)$ gauge fields. One needs a prescription that specifies how this gauge field interacts with the $2+1$ dimensional Dirac fermion. 
 The fermion coupling to the gauge fields can be based on $2+1$ dimensional gauge invariance since that will necessarily preserve two dimensional gauge invariance. One then begins with a $2$ dimensional dynamical $U(1)$ gauge field fluctuating on the disk domain wall at $r=R$ and for each such gauge field configuration solves the $2+1$ dimensional equation of motion for $U(1)$ gauge fields, i.e. Maxwell's equations to obtain a corresponding $2+1$ dimensional gauge field configuration. Additionally, in this work we demand that the gauge configuration should be a stable local minimum of the $2+1$ dimensional gauge action (Maxwell action in this case).
 This specifies the full gauge-fermion coupling.  
 For non-Abelian gauge fields in $2n$ dimensions, one must solve the $2n+1$ dimensional non-Abelian gauge field EOM (Yang-Mills equation of motion). For the example at hand, this prescription ensures that the gauge fields only propagate in the two dimensions $t, \phi$. 
 Since the fermion mode at $r=R$ for Eq. \ref{mr} is a Weyl fermion of negative chirality, we expect non-conservation of fermion current due to the chiral anomaly leading to
 \beq
\int_{R-\Delta}^{R+\Delta} dr\,\, \partial_\mu j^Q_\mu=\frac{Q E_\phi}{2\pi}
 \label{eq:anom}
 \eeq
 where $Q$ is the charge of chiral mode, $j^Q_{\mu}$ is the fermion current in the $2+1$ dimensional theory,  $\mu=t, \phi$, $E_\phi$ is the electric field in $\hat{\phi}$ direction at $r=R$  and $2\Delta$ is the support of the chiral mode around $r=R$ extending from $R-\Delta$ to $R+\Delta$.  We will take $\Delta$ to be the region in which $98\%$ of the chiral mode charge density lies. Note that Eq. \ref{eq:anom} holds as long as $E_\phi$ is held constant over this support range. If the cutoff $m_0$ is taken to infinity, $\Delta$ reduce to zero. On the other hand, when it is finite but large, $\Delta$ is small, but nonzero. For a mode of opposite chirality the sign on the RHS of Eq. \ref{eq:anom} changes. Note that $j$ itself is a $2+1$ dimensional current that is exactly conserved, which immediately implies that the RHS of Eq. \ref{eq:anom} is related to the radial component of the current $j_r$ via

  \beq
 \int_{R-\Delta}^{R+\Delta} dr\,\, \frac{1}{r}\partial_r(r j^Q_r) =-\frac{Q E_\phi}{2\pi}.
 \label{eq:anom2}
 \eeq
 
 This is known as anomaly inflow. Note that, in the $3-4-5-0$ model, one finds that the weighted sum of 
$\sum_Q Q\int_{R-\Delta}^{R+\Delta} dr\,\, \partial_\mu j^Q_\mu=0$. 
 In this paper, our aim is to translate the gauge field prescription of \cite{Kaplan:2024ezz} on a square lattice and demonstrate how the prescription works using the example of one specific two dimensional gauge field configuration. We then couple this gauge field to a charge $Q=1$ Weyl fermion to show anomaly inflow on the lattice. Subsequently, we couple the flow gauge field to the $3-4-5-0$ fermions and demonstrate lattice anomaly cancellation.  

 {\bf Chern-Simons current:} In general, it is difficult to compute $j^Q_i$ analytically for a manifold with a mass defect as is the case here. However, one could integrate out the gapped fermion modes for $r\ll R$ or $r\gg R$, which leads one to the expression of the current in terms of Chern-Simons functions given by
 \beq
j^{Q}_i=\begin{cases}\frac{Q}{4\pi}\epsilon_{ijk}\bar{F}_{jk}\,\,\,\,\text{for}\,\, r\ll R \\
0\,\,\,\,\,\,\,\,\,\,\,\,\,\,\,\,\,\,\,\,\,\,\,\,\,\,\text{for}\,\, r\gg R
\label{cs}
\end{cases}
\eeq
where $\bar{F}$ is the $2+1$ dimensional field strength tensor given by $\bar{F}_{ij}=\partial_i \bar{A}_j-\partial_j \bar{A}_i$ with $\bar{A}$ being the $2+1$ dimensional gauge field. Note that, 
according to our convention here, $\epsilon_{tsx}=1$.
The expression for the current in Eq. \ref{cs} starts breaking down near $r\approx R$. However, it will give us a good qualitative understanding of what to expect from a square lattice analysis later in this paper.

{\bf Gauge field:}
The target theory of interest here will have arbitrary $1+1$ dimensional gauge field configuration at $r=R$, for $A_\phi$ and $A_t$. With a completely localized boundary fermion mode at $r=R$, the prescription of \cite{Kaplan:2023pxd, Kaplan:2024ezz} demands that for $r<R$, $\bar{A}_\phi$ and $\bar{A}_t$ satisfy equation of motion, which in this case is Maxwell's equation
subject to the boundary condition:
\beq
\bar{A}_t(r=R)=A_t\nonumber\\
\bar{A}_\phi (r=R)=A_\phi.
\label{bc}
\eeq
Since we only care about $2$-dimensional gauge invariance, we can set $\bar{A}_r=0$ everywhere.  
In addition, we note that the equation of motion by itself is not sufficient in providing a unique gauge configuration. An additional condition is needed. In this paper, we provide this additional condition by demanding that the $2+1$ dimensional gauge field be a stable local minimum of the $2+1$ dimensional gauge field action whose minimization leads to the equation of motion, while holding $\bar{A}_r=0$.

There are further subtleties with the gauge field prescription which arise when 
the support for the boundary fermion mode has nonzero width, which we can denote as $2\Delta$, extending from $R-\Delta$ to $R+\Delta$. 
If we prescribe that the $\bar{A}_\mu$ gauge fields satisfy EOM for 
in $R>r>R-\Delta$, it will introduce $r$ dependence in the gauge fields within the support region, subjecting the boundary fermion to $2+1$ dimensional magnetic field. This is undesirable and we modify the gauge field prescription to avoid it. To obtain the right prescription, we first write $2+1$ dimensional  magnetic field in cylindrical coordinates:

\beq
\bar{B}&=&\frac{1}{r}\left(\frac{\partial (r \bar{A}_\phi)}{\partial r}-\frac{\partial \bar{A}_r}{\partial \phi}\right)\rightarrow \frac{1}{r}\left(\frac{\partial (r \bar{A}_\phi)}{\partial r}\right)
\label{eq:B}
\eeq
where the last line came from setting $\bar{A}_r=0$. Thus, we need 

\beq 
r \bar{A}_\phi(r)=C
\label{eq:raphi}
\eeq 
at least within $R+\Delta>r>R-\Delta$ in order to avoid the boundary fermions interacting with $2+1$ dimensional magnetic fields. At this stage we demand that Eq. \ref{eq:raphi} hold in the region $R+\Delta>r>R-\Delta-\delta$ with a small $\delta\neq 0$,  
with $C$ being $r$-independent to eliminate any magnetic field $B$ in the support region. 
Since the target gauge field config at $r=R$ is $A_t$ and $A_\phi$, we set
\beq
C=R A_\phi \implies \bar{A}_\phi(r)=R\frac{A_\phi}{r} \,\,\nonumber\\
\label{aphi}
\eeq
in $R+\Delta>r>R-\Delta-\delta$.
Furthermore, within the same region, we aim to avoid a $2+1$ dimensional radial electric field which can be expressed as 
\beq
\bar{E}_r=-\frac{\partial A_r}{\partial t}+\frac{\partial A_t}{\partial r} \rightarrow \frac{\partial A_t}{\partial r}.
\label{eq:Er}
\eeq
Thus, we maintain
\beq
\bar{A}_t=A_t
\label{at}
\eeq
in this region as well.
Eq. \ref{aphi} and \ref{at} together give us the gauge field prescription in $R+\Delta>r>R-\Delta-\delta$.

When specifying a gauge field prescription for 
$r < R-\Delta-\delta$
, we need to avoid multi-valued field strengths at $r\rightarrow 0$ whereas for $r> R$, there are no such restrictions. One of the ways to arrive at single-valued field strengths at $r\rightarrow 0$ is to solve 
equation of motion in the interior of the disk, i.e. 
$r < R-\Delta-\delta$.
The absence of 
similar restrictions for $r>R$, implies that 
we can impose Eq. \ref{aphi} and \ref{at} for any $r>R$.

In the region 
$r < R-\Delta-\delta$
, we will implement the equation of motion flow subjected to $\bar{A}_r=0$
with 
\begin{widetext}
\beq
    \frac{1}{r}\partial_\phi\partial_t \bar{A}_t-\partial_t^2 \bar{A}_\phi
    -
    \partial_r
    \left[
    \frac{1}{r}(\partial_r(r\bar{A}_\phi))
    \right]
    = 0 
    \label{eq:blk}
\eeq

and 
\beq
    \partial_r
    \left(
    r(\partial_r \bar{A}_t)
    \right)
    +
    \partial_\phi
    \left(
    \frac{1}{r}\partial_\phi \bar{A}_t-\partial_t \bar{A}_\phi
    \right)
     = 0 
     \label{eq:blk2}
\eeq
while selecting for the stable local minimum of the $2+1$ dimensional gauge action. 
Thus, for  
$r < R-\Delta-\delta$, 
$\bar{A}_t$ and $r\bar{A}_\phi$ will now become $r$ dependent in general in contrast with Eq. \ref{aphi} and Eq. \ref{at}. 

\end{widetext}

We conclude this continuum section by specializing to a boundary gauge field configuration that was considered before \cite{jansen} on a slab for regular domain wall fermions and CGT on a slab in \cite{PaperA}. Before we move to a lattice realization of the flow for general boundary gauge fields in the next section, our goal here is to get analytical results for this specific gauge field configuration in continuous space-time which we can compare with lattice results later. Translating the boundary gauge field configuration of \cite{jansen} and \cite{PaperA} for the disk at $r=R$ leads to
\beq
A_\phi= \frac{L_t}{2\pi}E_0 \cos(\omega t), \,\,A_t=0
\label{eq:ans}
\eeq
with $\omega=\frac{2\pi}{L_t}$ where $L_t$ is the length of $\mathbb{T}_1$ portion of the manifold.

Thus, according to our prescription, for
$r> R-\Delta-\delta$
we have 
\beq
\bar{A}_\phi=\frac{R A_\phi}{r}=\frac{RL_t}{2\pi r}E_0 \cos(\omega t).
\label{eq:innbdry}
\eeq

The solution to the EOM gauge field in 
$r < R-\Delta-\delta$
can be obtained using the ansatz $f(t)\mathcal{R}(r)$
with 
\beq
f(t)=\cos(\omega t)
\eeq
and

\begin{equation}
    \R(r) = c_1 I_1(\lambda r) + c_2 K_1(\lambda r)
    \label{eq:calR2}
\end{equation}
where
$I_1$, and $K_1$ are modified Bessel's function of the first and the second kind. The first kind does not diverge at $r=0$, but the second does. Since we are aiming to obtain the stable local minimum of the $2+1$ dimensional gauge field action, we set $c_2=0$ to obtain for 
$r < R-\Delta-\delta$,
\beq
    &&\bar{A}_\phi(t,\phi, r)\nonumber\\
    &=& \frac{R}{R-\Delta} E_0 \frac{L_t}{2\pi}\cos{\left(\frac{2\pi}{L_t}t\right)} \frac{I_1\left(\frac{2\pi }{L_t}r\right)}{I_1\left(\frac{2\pi }{L_t}(R-\Delta)\right)}\nonumber\\
    \label{eq:Aphi2}
\eeq
where we have used Eq. \ref{eq:innbdry} as a boundary condition at 
$r= R-\Delta-\delta$.
The solution for $\bar{A}_t$ in 
$r < R-\Delta-\delta$
is trivial and is given by
\beq
\bar{A}_t(t, \phi, r)=0
\label{eq:at2}
\eeq

We can now compute the corresponding electric and magnetic fields for $r<R-\Delta-\delta$
\begin{equation}
    \begin{aligned}
        &\vec{E}  = \tilde{C}\sin(\omega t) \omega  I_1(\omega r) \hat{\phi}\\
        &B  = \frac{1}{r}\tilde{C}\cos(\omega t) \left( \omega r I_2(\omega r) + 2 I_1(\omega r)  \right) \label{eq:EBsolutionEoM}
    \end{aligned}
\end{equation} 
where 
\beq
\tilde{C}=\frac{R}{R-\Delta} E_0 \frac{L_t}{2\pi} \frac{1}{I_1\left(\frac{2\pi }{L_t}(R-\Delta)\right)}.
\eeq
For 
$r > R-\Delta-\delta$
we simply have:
\begin{equation}
    \begin{aligned}
        &\vec{E} = \frac{R}{r}E_0 \sin(\omega t) \hat{\phi}\\
        &B  = 0 \label{eq:CynlindricalEB}
    \end{aligned}
\end{equation}
as set by the condition Eq. \ref{eq:innbdry}. 
Thus, we can now substitute the gauge fields in the expression for the current density in Eq. \ref{cs} to obtain its behavior for all $r$, albeit qualitatively so near $r\approx R$. Thus, for 
$r < R-\Delta-\delta$
we find 
\begin{equation}
    \begin{aligned}
        &j^t = \frac{q}{4\pi} \frac{1}{r}\tilde{C}\cos(\omega t) \left( \omega r I_2(\omega r) + 2 I_1(\omega r)  \right)  \\
        &j^r = \frac{q}{4\pi} \tilde{C} \sin(\omega t) \omega I_1(\omega r) \\
        &j^\phi = 0 \label{eq:CScurrentInDisk}
    \end{aligned}
\end{equation}
whereas for $r>R+\Delta$
\begin{equation}
    \begin{aligned}
        &j^t = 0  \\
        &j^r = 0  \\
        &j^\phi = 0. \label{eq:CScurrentsOutDisk}
    \end{aligned}
\end{equation}

Note that the expression for the current in Eq. \ref{eq:CScurrentInDisk} peaks near 
$r\sim R-\Delta - \delta$.
While it is true that the expression starts breaking down at this point, the qualitative behavior captured by the expression is correct. This is to say, we expect the bulk current density to peak near 
$r\sim R-\Delta - \delta$
even if they don't follow the exact expression in Eq. \ref{eq:CScurrentInDisk}. 
\section{Lattice Realization}
\label{sec:latimpl}
We will now build a prescription for how one can translate the continuum flow described in the previous section to a square lattice for a general boundary gauge field configuration. 

\begin{figure}
    \centering
\includegraphics[width=0.4\textwidth]{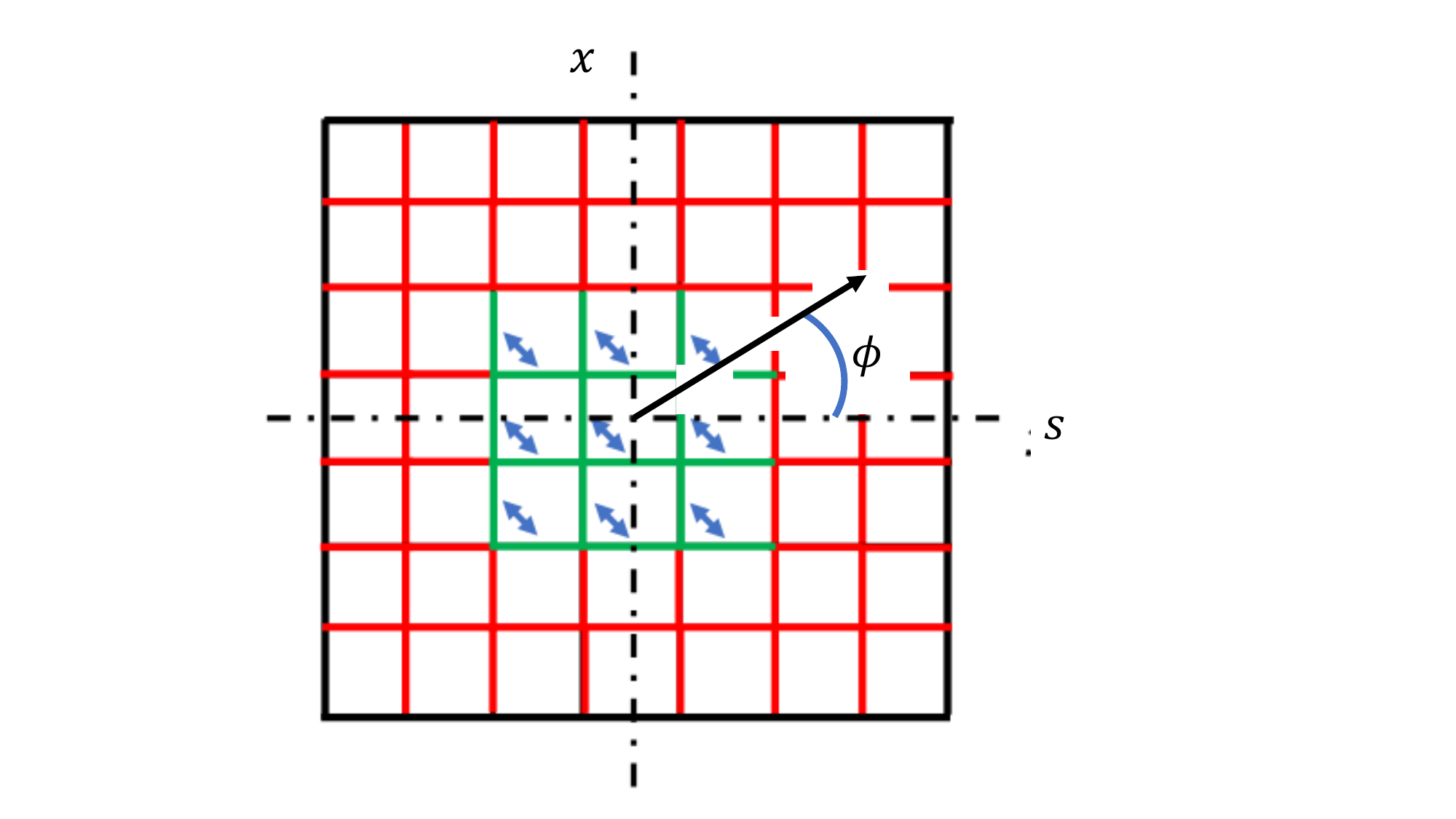}
    \caption{Lattice gauge field prescription: we show $8\times 8$ lattice here. The origin is taken at where the two dashed black lines denoting $s$ and $x$ axes intersect. The solid black lines show the location of the mass defect, i.e. the sign of fermion mass changes between inside the black boundary and outside. The gauge links on the red links follow Eq. \ref{aphi} and Eq. \ref{at}. The gauge links on the green links follow equation of motion, i.e. the lattice analog of Eq. \ref{eq:blk} and Eq. \ref{eq:blk2}. The green links on the two ends of the left-right arrows are related to each other via Eq. \ref{eq:axay}.}
    \label{disk-lat2}
\end{figure}

\noindent
{\bf Illustration of the flow in a simplified setting:}
 As a first step we illustrate the flow in a simplified setting. The goal of this discussion is to demarcate which links follow the flow equations and which do not without going into how exactly the flow is implemented. In the subsequent discussion, we will illustrate the flow implementation. 
 The simplified setting for this discussion deforms the disk into a square. Thus the domain wall instead of being circular has a square boundary. Moreover, for this illustration we will focus only on the interior of the domain wall. This is because as mentioned earlier, the gauge field prescription in the exterior region of the disk $r>R$ is similar to the prescription for the outer regions of the interior of the disk, i.e.  
 $R>r>R-\Delta-\delta$.
 The same will hold for a square boundary. 
We will illustrate the prescription using an $L_s=8, L_x=8$ lattice in Fig. \ref{disk-lat2}. In fig \ref{disk-lat2} we are showing the ($s, x$) plane of the lattice. The black boundary in the figure corresponds to the location of the domain wall in fermion mass. In the interior of it the fermion mass is positive whereas it is negative outside. We pick a coordinate system where $s=0, x=0$ is not located on any lattice site. Instead, it is at the center of the central plaquette as shown in Fig. \ref{disk-lat2} at the intersection of the dashed line, the $s$ and $x$ axes. 

Every link on this lattice is associated with a corresponding gauge link. We denote the coordinate associated with the gauge link  along $i$ where $i\in s,x,t$ as $n$.
It is given by the coordinate of the lattice site on one of its ends with the smaller $n_i$ value. 
Again, to simplify the illustration, we deform the boundary of the region in which we intend to solve the EOM, into a square as well. We take this region to be an inner square of lattice size $(2l+1)\times (2l+1)$. For the links outside we want to implement Eq. \ref{aphi} and Eq. \ref{at}. The interior region square boundary extends from $l-1/2>s, x>-l-1/2$. The outer region where we implement Eq. \ref{aphi} and Eq. \ref{at} is $s>l-1/2$, $s<-l-1/2$, $x>l-1/2$ and $x<-l-1/2$.

We identify the $s,x$ coordinates of each lattice site in the region $x>l-1/2$, $x<-l-1/2$, $s>l-1/2$ and $s<-l-1/2$, and solve for the corresponding azimuthal coordinate $\phi=\tan^{-1}(x/s)$. Thus, the temporal gauge links attached to the lattice site can be obtained by substituting $(t, \phi)$
values of the corresponding site in the function $A_t$. In other words, we replace the gauge link $U_t$ by the the function $e^{i A_t}$ evaluated at $\phi$ and $t$ associated with the lattice site. 
$A_s$ and $A_x$ or $U_s$ and $U_x$ can be obtained similarly, by evaluating the function $\frac{A_\phi}{r}$ at the location of the appropriate lattice site.

 For every link $U_s(s, x, t)$, $U_x(s, x, t)$ and $U_t(s, x, t)$ within $-l-\frac{1}{2}<x<l-\frac{1}{2}$ and $-l-\frac{1}{2}<y<l-\frac{1}{2}$, one solves the EOM. In Fig. \ref{disk-lat2}, $l=1$. Thus, all the green links, as well as gauge links along $t$, $U_t$ within this region(not shown in the figure) are obtained by solving EOM. Furthermore, all the green links are not independent since $\bar{A}_r=0$. This will relate $A_s(s,x,t)$ and $A_x(s,x, t)$ leaving only one independent variable from the two. 
The relation between $A_s$ and $A_x$ can be obtained from 
\beq
A_s\cos\phi + A_x\sin\phi =0\nonumber\\
\implies U_s^{\cos\phi}=U_x^{-\sin\phi}.
\label{eq:axay}
\eeq
 The doubly directed arrows in the figure Fig. \ref{disk-lat2} shows which two links are related to each other. 
 To implement the EOM for the links $-l-\frac{1}{2}<s<l-\frac{1}{2}$ and $-l-\frac{1}{2}<x<l-\frac{1}{2}$, we write down the plaquette action 
\begin{equation}
    \hat{S} = \sum_{n}\sum_{i,j}\mathrm{Re}\ \mathrm{tr}\{ 1 - U_\mathrm{plaquette}(n,i,j) \}. 
    \label{hats}
\end{equation}
where $n$ goes over all lattice sites and $i, j$ take values in $s,x,t$ with  
\beq
U_\mathrm{plaquette}(n,i,j)
&=& U_i(n,i)U_j(n+\hat{i},j)\nonumber\\
&&U_i(n+\hat{j})^{-1}U(n)_j^{-1}. 
\eeq
Minimizing this action with respect to the green links along $s$ and $x$, i.e. $U_s$ and $U_x$ and $U_t$ for $-l-1/2<x,s<l-1/2$ while imposing Eq. \ref{eq:axay} will yield the flow gauge links. 

\subsection{Implementation of the minimization using additional gradient flow}
A simple way to execute the equation of motion flow for the gauge field in the interior region is to add an extra parameter direction $\tau$ and allow the gauge field $\bar{A}$ to depend on $\tau$ along with the coordinates $t, s,x$ or $t, r, \phi$. In the continuum language we then
evolve the gauge field as a function of $\tau$ within the interior according to the following equation:
\beq
\partial_\tau \bar{A}_\mu=\frac{\xi}{|\Lambda|}\partial_i \bar{F}_{i\mu}
\label{eq:gr}
\eeq
where $\mu, \nu$ subscript stand for $t, \phi$ whereas the subscript $i$ stands for $t, \phi, r$, while holding the gauge field in the outer region fixed to Eq. \ref{aphi} and \ref{at}. $\xi$ is a number and we set to $1$, even though its precise value is  unimportant for this discussion.  $\Lambda$ is the cutoff scale which on the lattice can be set to the inverse lattice spacing. 
To solve Eq. \ref{eq:gr}, we will need an initial condition at $\tau=0$. To set this condition, we can allow the gauge field at $\tau=0$ to satisfy Eq, \ref{aphi} and \ref{at} everywhere.  
After long enough time $\tau$, the right hand side of Eq. \ref{eq:gr} equals zero, i.e. the
gauge fields satisfy equations of motion:
\beq
\partial_i \bar{F}_{i\mu}=0
\label{eq:fl}
\eeq
, in other words, Eq. \ref{eq:blk} and \ref{eq:blk2}. One of the benefits of the gradient flow in $\tau$ is that it not only leads us to a solution to 
the equation of motion, it also naturally selects the stable local minimum of the gauge action as desired.\\ 

\noindent
{\bf Lattice version of the additional gradient flow:}
We want to make the above prescription work on the lattice. We begin with the relation between $U_r, U_\phi$ and $U_s$, $U_x$
\begin{equation}
    \begin{aligned}
        &U_r(n) = \left[ U_s(n) \right]^{\cos\phi} \left[ U_x(n) \right]^{\sin\phi}, \\
        &U_\phi(n) = \left[ U_s(n) \right]^{-\sin\phi} \left[ U_x(n) \right]^{\cos\phi}. \\
    \end{aligned}
\end{equation}

\begin{equation}
    \begin{aligned}
        &U_s(n) = \left[ U_r(n) \right]^{\cos\phi} \left[ U_\phi(n) \right]^{-\sin\phi}, \\
        &U_x(n) = \left[ U_r(n) \right]^{\sin\phi} \left[ U_\phi(n) \right]^{\cos\phi}. \\
    \end{aligned}
\end{equation}

Then, for the Lie group derivatives defined in \cite{luscher2010trivializing}, for a functional $f$ of the gauge links, we have
\beq
    \partial_{n,\phi}f[U] \equiv \left. \frac{d}{d\kappa}f[U_\phi(n)] \right|_{\kappa=0},
    \eeq
    with
    \beq
    \frac{dU_\nu(\tilde{n})}{d\kappa} = \left\{
    \begin{array}{cc}
        e^{-i\kappa\sin\phi}U_s(n) & \text{, if } (\tilde{n},\nu) = (n,s) \\
        e^{+i\kappa\cos\phi}U_x(n) & \text{, if } (\tilde{n},\nu) = (n,x) \\
        U_\nu(\tilde{n}) & \text{else.}  \\
    \end{array}\right.
\eeq
So, it is easy to find that
\begin{equation}
    \partial_{n,\phi} f[U] = -\sin\phi \partial_{n,s} f[U] + \cos\phi \partial_{n,x} f[U].
\end{equation}
We define
\[ \hat S = \sum_n\sum_{\mu,\nu}\mathrm{Re}\{1-U_{\mathrm{plaquette}}(n,\mu,\nu)\}\]
such that
\begin{equation}
     Z_{n,\mu}[U] = - \partial_{n,\mu}\hat S.
\end{equation}
Here, note that $Z_{n,\mu}[U] \in U(1) = \{ i\theta|\theta\in\mathbb{R}\}$. In polar coordinates this leads to
\begin{equation}
    Z_{n,\phi} \equiv -\partial_{n,\phi} \hat S= -\sin\phi Z_{n,s} + \cos\phi Z_{n,x}
\end{equation}
and the corresponding gradient flow for $U_\phi(n)$ is
\begin{equation}
    \partial_\tau U_\phi(n) = -\xi\left\{ \partial_{n,\phi} \hat S \right\} U_\phi(n).
    \label{eq:up}
\end{equation}
It is now simple to obtain the numeric iteration algorithm where the updated links are denoted with a primed superscript, given by
\begin{equation}
    U_\phi'(n) = \exp(\xi\epsilon Z_{n,\phi})U_\phi(n)
\end{equation}
which in conjunction with $U_r'=1$ implies
\begin{equation}
    \begin{aligned}
        U_s'(n) = \exp(-\xi\epsilon Z_{n,\phi}\sin\phi)U_s(n) \\
        U_x'(n) = \exp(+\xi\epsilon Z_{n,\phi}\cos\phi)U_x(n). \label{eq:GFDisksc}
    \end{aligned}
\end{equation}

In general, one is to implement a gradient flow for the time component of the gauge link $U_t$ as well. However, since we know that our specific boundary gauge configuration does not result in a radial dependence for the time component, we refrain from updating it.

\section{Flow gauge field and fermions}
\label{sec:ferm}
We previously used a lattice (Fig. \ref{disk-lat2}) with square boundaries to illustrate the idea of flow. We will now return to the original setup with a square lattice which contains a disk defect for the fermion mass. In this analysis we will retain the exterior of the disk which we had omitted for convenience in the illustration involving the simplified square boundary.

 We will first discuss the fermion sector, which is then followed by the gauge fields. The fermion action (charge $1$) coupled to gauge fields has the form

\beq
        S &=& \frac{1}{2}\sum_{n,i} \left( \bar{\psi}_{n} \gamma_i U_{i}(n) \psi_{n+\hat{i}} - 
    \bar{\psi}_{n+\hat{i}} \gamma_i U_i^*(n) \psi_n \right)\nonumber\\
    &+& m_0\sum_n \bar{\psi}_n \psi_n \nonumber\\
    &+& w \sum_{n,i} \left( -2\bar{\psi}_n \psi_n + \bar{\psi}_n U_{i}(n) \psi_{n+\hat{i}} +
    \bar{\psi}_{n+\hat{i}} U_i^*(n) \psi_n \right)\nonumber\\   \label{eq:action}
    \eeq
where $i \in \{ t,s, x \}$, $w$ is the Wilson parameter, and
\begin{equation}
    n \in \Lambda = \{(t,s,x)| t \in  \mathbb{Z}_{L_t}, s \in \mathbb{Z}_{L_s},x \in \mathbb{Z}_{L_x} \}.
\end{equation}
We define the radial and azimuthal coordinate on the lattice as
\begin{equation}
   \begin{aligned}
        &r = \sqrt{(s-\lfloor L_s/2 \rfloor + 1/2)^2+(x-\lfloor L_x/2 \rfloor + 1/2)^2},\\ 
        &\phi = \arctan{\frac{x-\lfloor L_x/2 \rfloor + 1/2}{s-\lfloor L_s/2 \rfloor + 1/2}} 
        \label{eq:Rphi}
   \end{aligned}
\end{equation}
and the domain wall mass profile is taken to be 
\begin{equation}
    m_0(r) \equiv \sinh(\mu_0)\theta\left(\frac{r}{R}\right) =  \begin{cases}
        -\sinh(\mu_0) & r \geq R, \\
        0, & R - 1 \leq r < R, \\
        +\sinh(\mu_0) & r < R,
    \end{cases}
    \label{m0}
\end{equation}
where $ R = \min\{\lfloor L_s/2 \rfloor,\lfloor L_x/2 \rfloor \}-1$ ( $\lfloor x\rfloor$ indicates floor of $x$). We set $\mu_0=0.972, w=0.566$ which leads to $m_0(r<R)=-m_0(r>R)=1.132$. 
The fermion-gauge action of Eq. \ref{eq:action} can be written in the form of $S = \sum_{nm} \bar{\psi}_n D_{nm} \psi_m$ with $D_{nm}$ being the entries of the fermion operator. 
Now, let's impose open boundary condition for ${x,s}$-direction and an anti-periodic boundary condition for ${t}$-direction on the fermion operator. We set $L_x=28, L_t=16, L_s=28$. Furthermore, we take $R=13$. The support of the fermions $\Delta\approx 4$.

With this choice we can now focus on the edge states of the system. For this, we turn off gauge fields by setting $U=1$ for all links. With the domain wall defect of Eq \ref{m0}, we get left chirality edge modes at $r=R$ in congruence with \cite{Kaplan:2023pvd}. Note that, open boundary in $x, s$ can in principle give rise to chiral modes localized at the boundary: i.e. $x=L_x$, $s=L_s$, $x=-L_x$ and $s=-L_s$, e.g. when $(m_0(r)/w)=1$ for $r>R$. Such modes would be undesirable. However, the profile in Eq \ref{m0} ensures that this condition is not satisfied and instead  $(m_0(r)/w)=-1$ for $r>R$, thus eliminating any undesirable extra chiral mode.

\begin{figure}
    \centering
\includegraphics[width=0.45\textwidth]{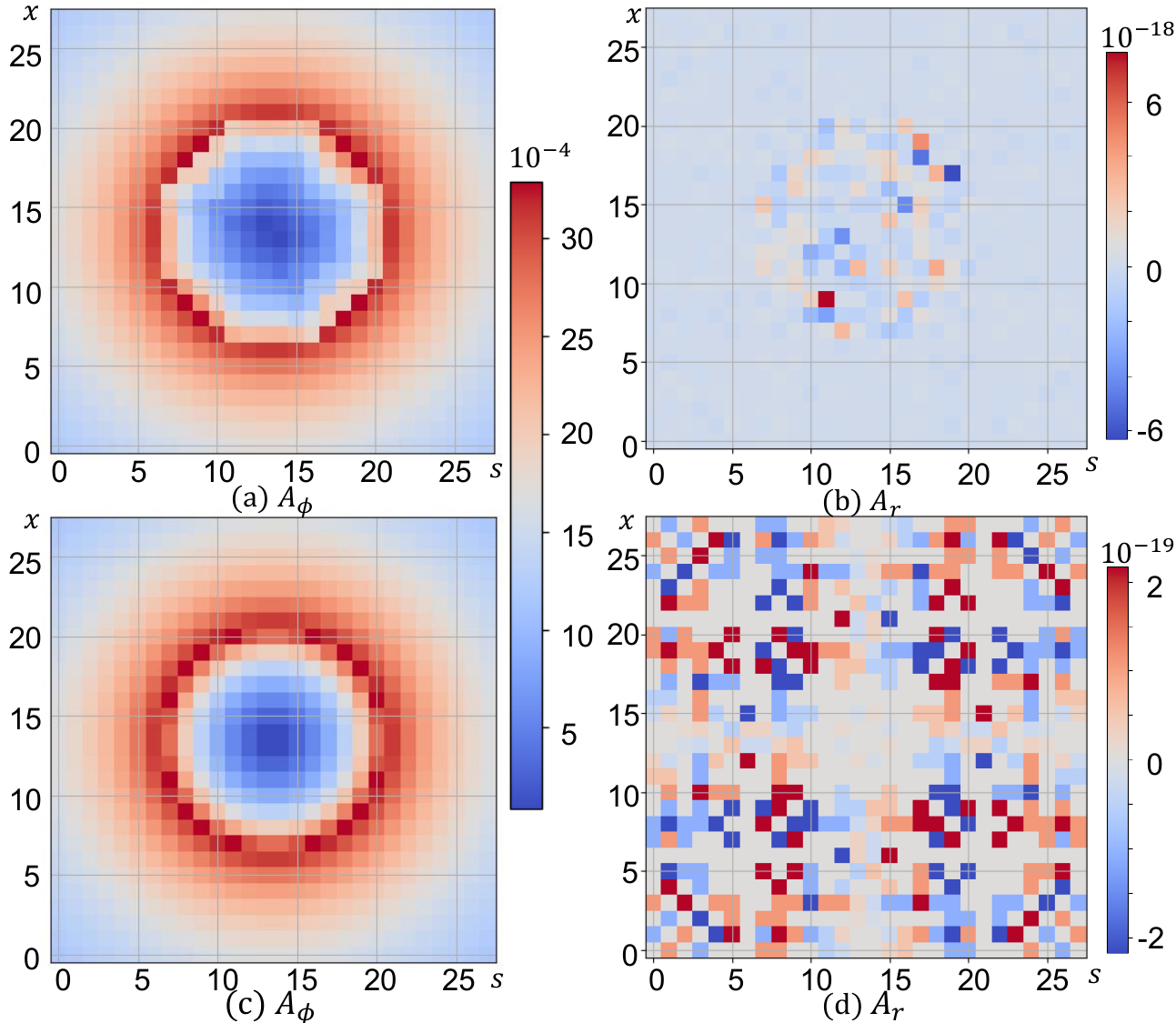}
    \caption{Gauge fields at $t=2$. Fig. (a,b) are the numerical flow gauge field results and Fig. (c,d) are the analytical solution as in Eq. \ref{eq:Aphi2}. }
    \label{fig:Aguage}
\end{figure}

\begin{figure}
    \centering
\includegraphics[width=0.45\textwidth]{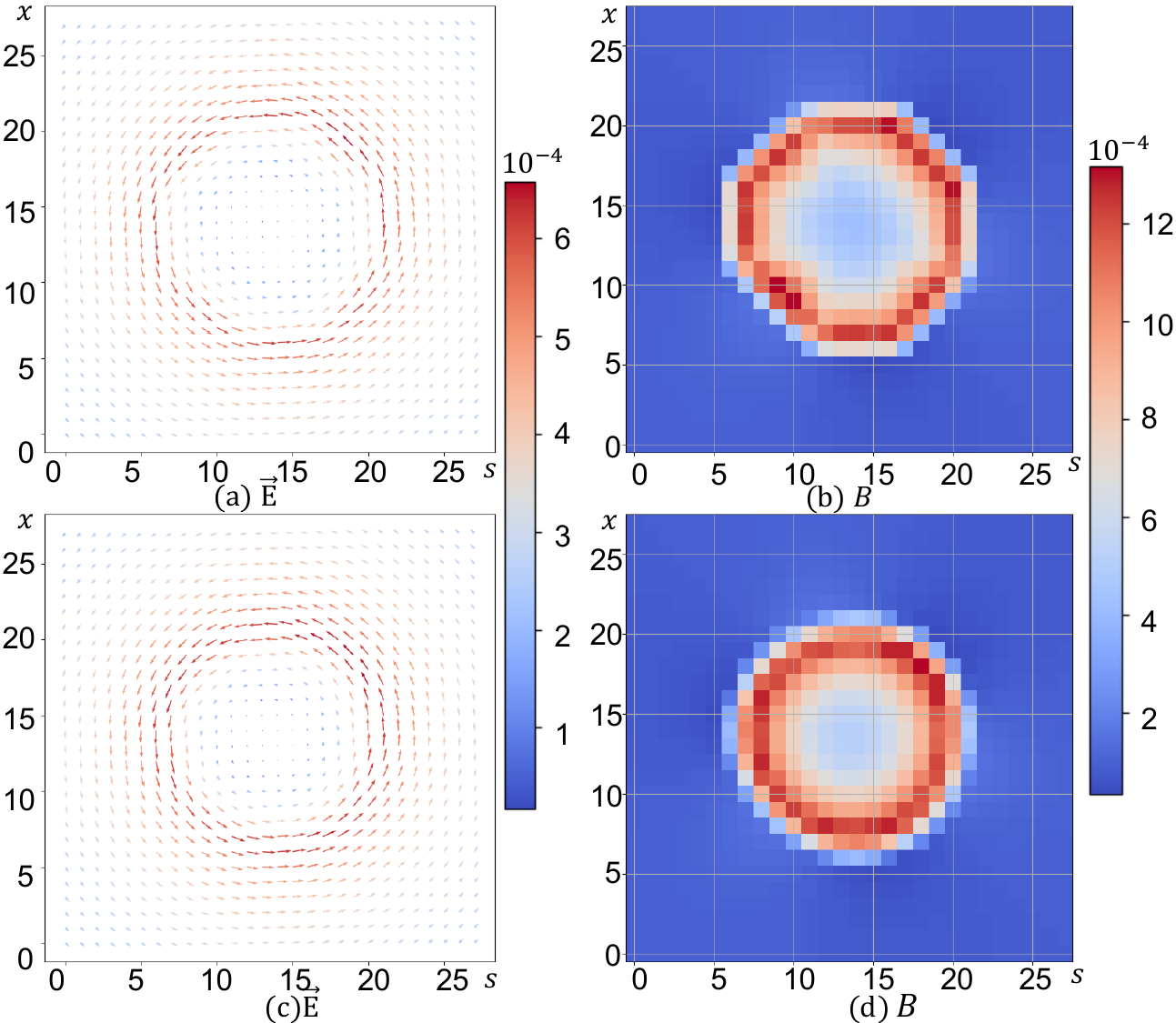}
    \caption{Electromagnetic fields at $t=2$. Fig. (a,b) are numerical results obtained from gauge field flow and Fig. (c,d) are the analytical solution as in Eq. \ref{eq:EBsolutionEoM}.  }
    \label{fig:EBfield}
\end{figure}

With the left chiral mode at $r=R$, we can turn on the gauge fields. From chiral anomaly, we expect the chiral mode to exhibit non-conservation of current according to the anomaly equation Eq. \ref{eq:anom}.
We will implement the flow for the specific boundary gauge field configuration of mimic Eq. \ref{eq:ans}. Just as discussed in section \ref{sec:ov}, we wish the gauge field prescribed as in Eq. \ref{eq:innbdry} for $r>R-\Delta-\delta$. We set $\delta=2$ and $E_0=0.001$, and solve for EOM on the links within $r<R-\Delta-\delta$ as suggested in the previous section \ref{sec:latimpl}, which involves introducing an additional imaginary time direction $\tau$ and implementing a gradient flow. For an initial condition at $\tau=0$, we use Eq. \ref{aphi}  and Eq. \ref{at} everywhere. 
We plot the numerical values for the gauge field after the flow has been completed, in the upper panel of Fig. \ref{fig:Aguage} for $t=2$. In the lower panel we plot the analytical results from Eq. \ref{eq:Aphi2} for the same time slice. We also plot the electric and magnetic fields in Fig. \ref{fig:EBfield}. On the left panel, we plot the numerical results at $t=2$ and on the right panel the analytical results for the same time slice using Eq \ref{eq:EBsolutionEoM}. As we can see, the numerical results are in good agreement with the analytical formula. We don't expect perfect agreement since the analytical expressions were derived in the continuum using derivative operators as opposed to finite difference operators.


We can now shift our focus on the current density. 
The current density operator can be extracted from the action in Eq. \ref{eq:action} and is given by
\beq
   && j_i = \frac{1}{2} [ \bar{\psi}_n \gamma_i [U_{i}(n)] \psi_{n+\hat{i}} +
    \bar{\psi}_{n+\hat{i}} \gamma_i [U_i^*(n)] \psi_n ] \nonumber\\
    &&
    + w [ \bar{\psi}_n  [U_{i}(n)] \psi_{n+\hat{i}} - 
    \bar{\psi}_{n+\hat{i}} [U^*_{i}(n)] \psi_n]. 
    \label{curr}
\eeq
We can now compute the current using 
\begin{equation}
    \langle j_{i} \rangle = \frac{1}{\hat{Z}} \int [D\bar{\psi} D\psi] e^{-S} j_{i} = \mathrm{Tr}[\hat{j}_{i} h^{-1}]
\end{equation}
where $h$ is the fermion operator extracted from Eq. \ref{eq:action}, $\hat{j}$ is the one body operator corresponding to the current density in Eq. \ref{curr} . $\hat{Z}$ is the partition function given by
\beq
\hat{Z}=\int [D\bar{\psi} D\psi] e^{-S}.  
\eeq

\begin{figure}
    \centering
\includegraphics[width=0.45\textwidth]{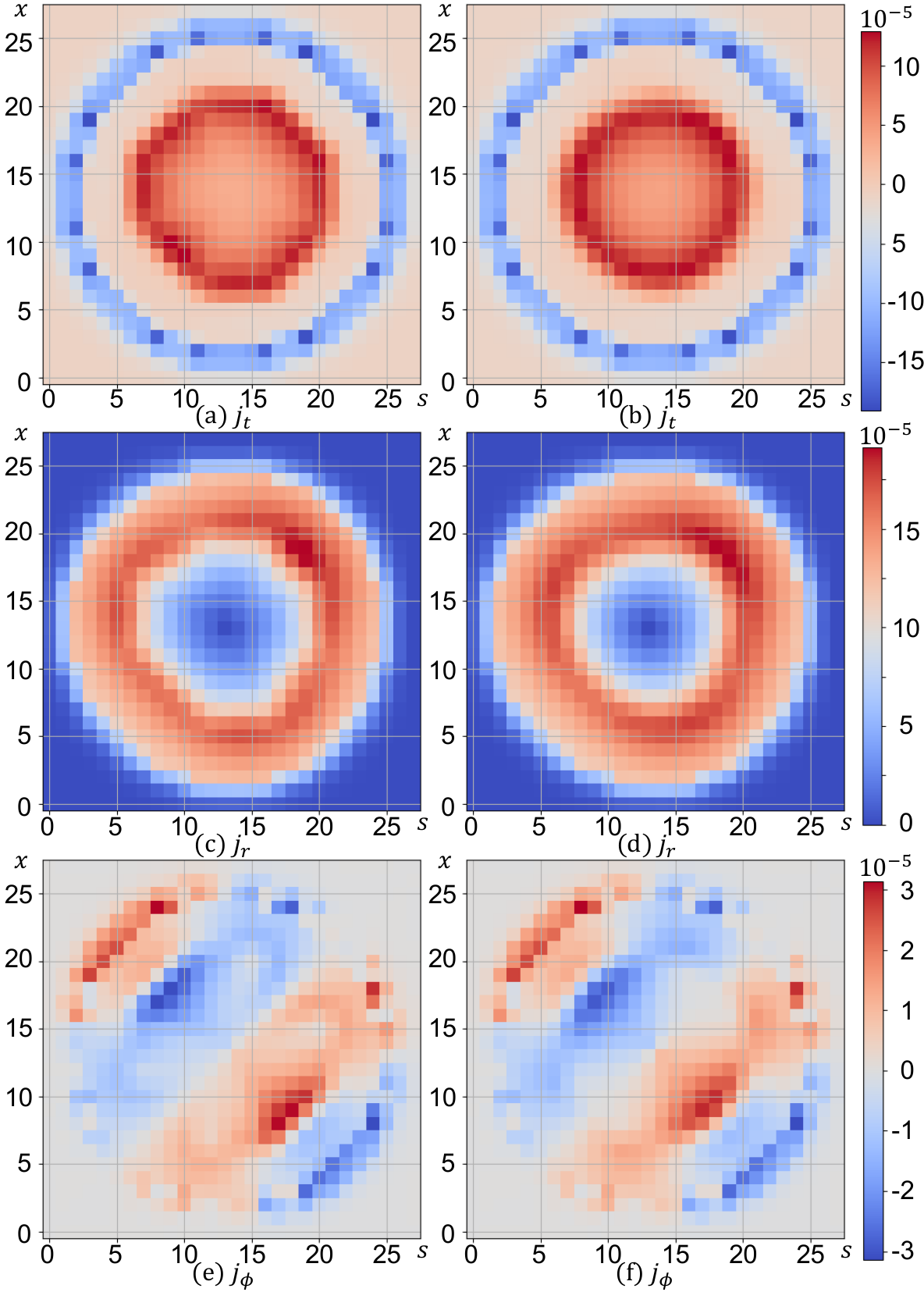}
    \caption{Currents at $t=2$. Fig. (a,c,e) are numerical results and Fig. (b,d,f) are from the gauge field analytical solution. }
    \label{fig:currents}
\end{figure}

In Fig. \ref{fig:currents} we plot $j_{r,\phi, t}$ at the time slice $t=2$ as a function of $s$ and $x$. On the left panel the currents are computed with the fermion operator interacting with the flow gauge field computed numerically using the prescription provided in section \ref{sec:latimpl}. The right panel shows the currents when fermion operator interacts with the analytical gauge field computed in Eq. \ref{eq:Aphi2} and \ref{eq:at2}. We again find good agreement between the results computed using numerically flowed gauge fields and analytically computed ones demonstrating that anomaly inflow works as expected on the lattice. An interesting feature of the results can be spotted from the plots for $j_t$ in this figure: one can see two concentric rings in the $s-x$ place, one colored blue, the other brown. The extent of the blue ring indicates the region over which the chiral fermion charge density peaks whereas the brown ring indicates the region where the bulk charge density due to the Chern-Simons operator peaks. The two regions are clearly separated, which will allow us to define a total charge for the boundary and a total charge in the interior later on.

On a square lattice, we can compute 
\beq
   \langle \partial_i j_j \rangle  = \langle j_j(n) - j_j(n-\hat i )\rangle ,
\eeq
where $i,j\in\{t,x,s\}$. So, effectively we define 

\beq
    j_r = j_s\cos\phi + j_x\sin\phi,\quad j_\phi = -j_s\sin\phi + j_x\cos\phi,
\eeq
and
\beq
    \frac{1}{r}\langle \partial_\phi j_\phi \rangle &=& -\frac{1}{r}\left(\cos\phi \langle j_s \rangle + \sin\phi \langle j_x \rangle \right)\nonumber\\ &&- \sin\phi\cos\phi \langle \partial_s j_x \rangle - \sin\phi\cos\phi \langle \partial_x j_s \rangle\nonumber\\ &&  + \sin^2\phi \langle \partial_s j_s \rangle + \cos^2\phi \langle \partial_x j_x \rangle .
\eeq
We now want to construct a two dimensional divergence of the $2+1$ dimensional current density in order to test Eq. \ref{eq:anom} on the lattice.
This means we have to sum the $2+1$ dimensional current density in small region around the domain wall at $r=R$ as in Eq. \ref{eq:anom} from $R-\Delta$ to $R+\Delta$. In addition, we choose to average over $\phi$ since for our specific choice of boundary gauge field, and the resulting currents, there is no $\phi$ dependence. Defining 
$\Lambda_\psi \equiv \{ (s,x) | R-\Delta  \leq r \leq R+\Delta \} $
, we can write
\beq
   \sum_{\Lambda_\psi} \langle \nabla \cdot \vec j \rangle \equiv  \sum_{\Lambda_\psi} \langle \partial_t j_t \rangle
   + \sum_{\Lambda_\psi} \langle  \frac{1}{r}\partial_\phi j_\phi \rangle .
\eeq 
To verify Eq. \ref{eq:anom} we define 
a ratio $\mathcal{R}$ as
\beq
    \mathcal{R} = \frac{ (Q E_{\mathrm{eff}}/2\pi)}{\frac{1}{2\pi R}\sum_{\Lambda_\psi}\langle \nabla\cdot\vec j \rangle},
\eeq
where $E_{\mathrm{eff}}(t) = E_0\frac{\sin(\omega)}{\omega}\sin(\omega t)$  with $Q=1$ for charge $1$ fermions. We expect this ratio to be close to $1$ if Eq. \ref{eq:anom} holds. For the numerically obtained flow gauge field, we find that $\mathcal{R}$ is  $1.023$ which is good agreement with the infinite volume result of $1$ as expected from Eq. \ref{eq:anom}. The slight deviation from $1$ is expected since $\mathcal{R}$ is necessarily being computed in finite volume. 

\begin{figure}
    \centering
\includegraphics[width=0.45\textwidth]{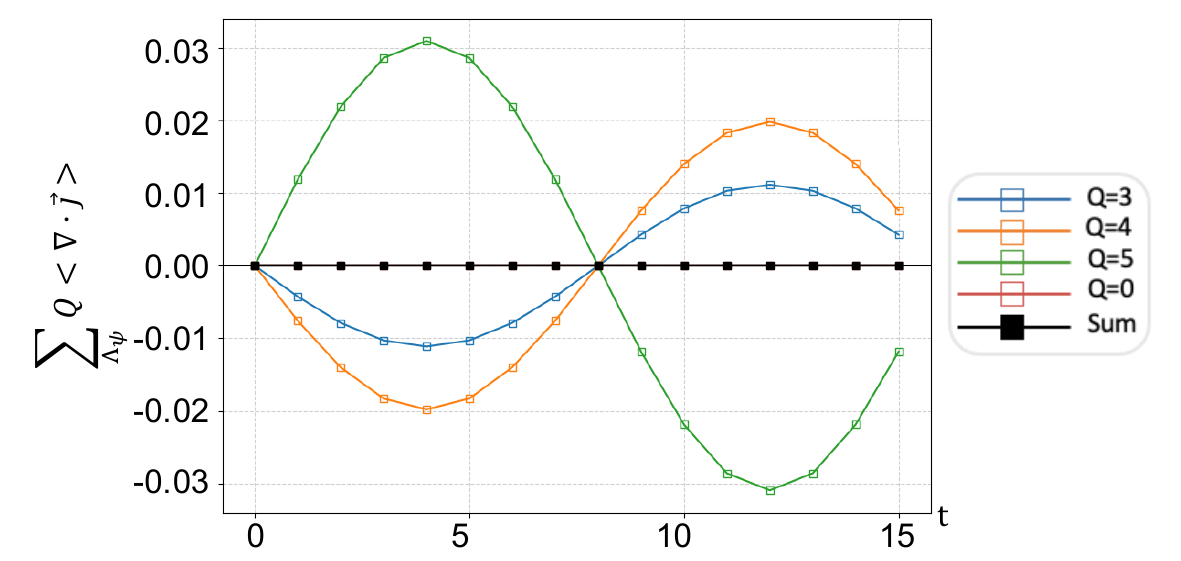}
    \caption{Two dimensional divergence summed over the region $\Lambda_\psi$, i.e.  $\sum_{\Lambda_\psi} Q <\nabla\cdot\vec j>$ for all flavors and their sum.}
    \label{fig:divj}
\end{figure}

It is straightforward to modify the above construction to engineer the $3-4-5-0$ model on the circular domain wall. 
We plot $\sum_{\Lambda_\psi} Q \langle \nabla \cdot \vec j^Q \rangle$ in Fig. \ref{fig:divj} as a function of time where $Q$ takes the values $3,4,5,0$ and 
$Q=3,4$ have the opposite chirality of $Q=5,0$. $j^Q$ denotes the current corresponding to the species with charge $Q$. We  show anomaly cancellation by summing up the currents with appropriate weight, i.e. $\sum_Q\sum_{\Lambda_\psi} Q \langle \nabla \cdot \vec j^Q \rangle =0$.

\begin{figure}
    \centering
\includegraphics[width=0.45\textwidth]{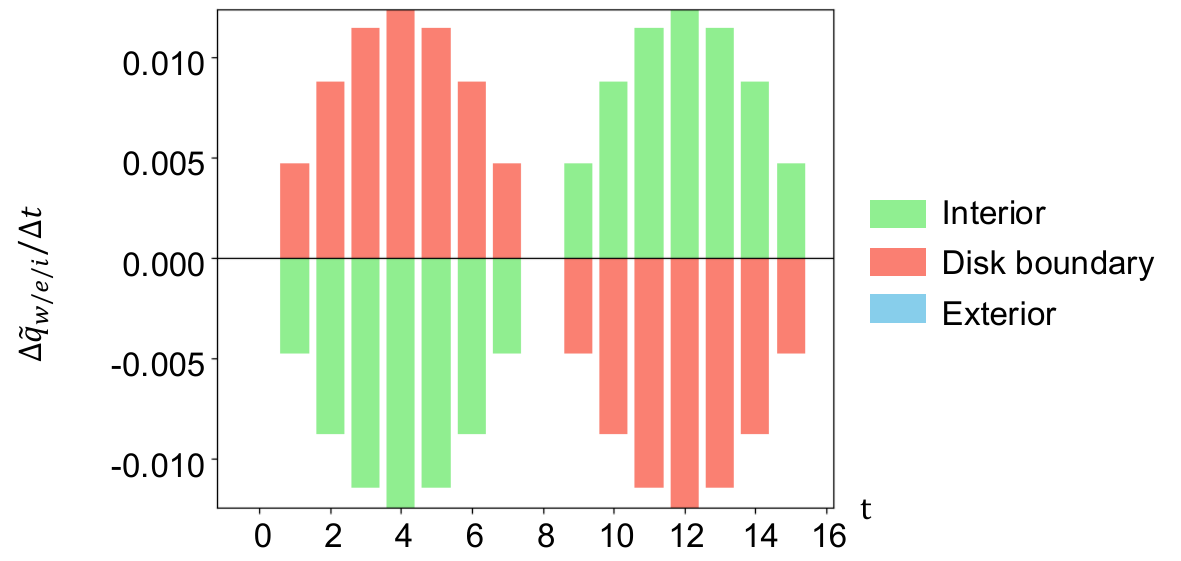}
    \caption{Rate of change of total charge in the interior, disk boundary and exterior.}
    \label{fig:DqDt}
\end{figure}

In addition, for the unit charge fermion, we define the total charge on the domain wall (disk boundary), interior region and the exterior as
\beq
    &&{q}_w(t) = \sum_{\Lambda_\psi} \langle j_t \rangle, \\
    &&{q}_i(t) = \sum_{\lambda_\psi} \langle j_t \rangle, \\
    &&{q}_e(t) = \sum_{\hat\lambda_\psi} \langle j_t \rangle,
\eeq
where
$\lambda_\psi = \{ (s,x) | r < R-\Delta \}$ 
representing the interior region of the disk and $\hat \lambda_\psi = \{ (s,x) | r > R+\Delta \} $ representing the exterior region of space outside the disk. We plot in
\ref{fig:DqDt} the rate of change of the net charge in these three region $\Delta q_{w,i,e}/\Delta t$ as a function of time where $\Delta t$ is taken to be $1$ in lattice units. We see that the non-conservation of charge on the boundary is compensated by the non-conservation of charge in the interior of the disk, maintaining overall charge conservation for the $2+1$ dimensional theory. Note that the exterior region does not participate in any exchange of charge with the boundary or the interior.

{\bf A comment on the modified fermion operator:} Note that, while we work with the fermion operator shown in the action $S$ in Eq. \ref{eq:action}, which we denote as $D$ where
\beq
S=\sum_{nm}\bar{\psi}_a D_{mn}\psi_n,
\eeq
\cite{Kaplan:2024ezz} proposed a slightly modified fermion operator for chiral gauge theory application. This modified operator is given by
\beq
\tilde{D}=\frac{D}{\sqrt{D^\dagger D+\mu^2\delta_{r,R}+\epsilon}}
\eeq
  where $\mu$ is some mass scale and $\delta_{r,R}$ is a delta function with support at $r=R$ and $\epsilon$ is a small scale with dimensions of mass squared to be taken to zero at the end of the calculation. $\tilde{D}$ has the same chiral mode spectrum as $D$ and as a result will have analogous behavior for anomaly inflow. For the purpose of this calculation it is not necessary to work with $\tilde{D}$ and hence we simply work with $D$. However in a simulation with dynamical gauge fields and quarks, one must use $\tilde{D}$ in order to eliminate the radiative corrections to the real part of the gauge field action coming from the fermion determinant.   

\section{Conclusion}
In this paper we have formulated the equation of motion flow prescribed in \cite{Kaplan:2023pxd, Kaplan:2024ezz} for the disk proposal on a square lattice for a $U(1)$ gauge field. In the continuum with arbitrarily localized chiral modes on the domain wall, the proposal involved specifying the gauge field configuration on the domain wall and then using higher dimensional equation of motion, in this case Maxwell's equation to determine the $2+1$ dimensional gauge field in the interior of the disk. On a finite lattice, with a chiral mode that has support over a region $2\Delta$, we  constrain the gauge field to avoid $2+1$ dimensional magnetic field and radial electric field over the support region. We implement the equation of motion flow in the interior by introducing an additional  imaginary time direction and using gradient flow in this direction which allows the interior gauge fields to reach the stable local minimum of the $2+1$ dimensional gauge action. Note that the flow prescription was implemented on a square lattice in its entirety. There may be alternative approaches. E.g. one may be able to formulate and implement a radial flow equation in polar coordinate and then use the results to obtain gauge links on a square lattice. The feasibility of such a direction is worth exploring in future work. 

We test our flow prescription for a specific gauge field configuration and find good agreement with the continuum analytical solutions of the higher dimensional equation of motion (the flow equation). We also analyze fermions currents in the background of this gauge field and successfully demonstrate the concept of anomaly inflow and anomaly cancellation working as expected on the lattice. Note that, our boundary gauge field configuration does not include a net nonzero instanton number. This allows us to avoid bulk singularities and corresponding zeromodes. These modes in the presence of net winding however, are an important ingredient of the formulation in \cite{Kaplan:2024ezz, Golterman:2024ccm, Aoki:2022aez} and should be explored in future work. Additionally, it is worthwhile exploring an annulus geometry with two concentric domain wall defects where the two walls host opposite chirality modes. Albeit, the modes living on the inner wall will have shorter wavelengths set by the inverse inner radius whereas the outer modes will have longer wave-lenths, set by the inverse outer radius.  

\section{Acknowledgement}
JD acknowledges the support of NSFC under Grants No. 12293060, No. 12293063.
RK is supported in part by the DOE NNSA LRGF Fellowship under cooperative agreement DE-NA0003960.
SS was supported by the U.S. Department of Energy,
Nuclear Physics Quantum Horizons program through the
Early Career Award DE-SC0021892.

\bibliography{ChiRef}

\end{document}